\documentclass[cits]{PoS}
\title{Search for Solar Axions with the CDMS-II Experiment}

\ShortTitle{Search for Solar Axions with the CDMS-II Experiment}
\author{\speaker{T. Bruch}\thanks{Physics Institute} , for the CDMS Collaboration\\
        University of Zurich, Winterthurerstrasse 190 8047 Zurich, CH\\
        E-mail: \email{tbruch@physik.uzh.ch}}

\abstract{The CDMS-II experiment operates 19 germanium detectors with a mass of 250g each in a very low background environment. Originally designed for the search for Dark Matter the experiment can also detect solar axions by Primakoff conversion to photons. The Bragg condition for X-ray momentum transfer in a crystal allows for coherent amplification of the Primakoff process. Since the orientation of the crystal lattice with respect to the Sun changes with daytime an unique pattern in time and energy of solar axion conversions is expected.

The low background $\sim$ 1.5 counts/kg/day/keV and knowledge of the exact orientation of all three crystal axes with respect to the Sun make the CDMS-II experiment very sensitive to solar axions. In contrast to helioscopes, the high mass region $<$ 1 keV can also be probed effectively. The alternating orientations of the individual crystals in the experimental setup provide different patterns of solar axion conversion, making a false positive result extremely unlikely.

The result of an analysis of 289  kg-days of exposure resulted in a null observation of solar axion conversion. This analysis sets an upper limit on the axion photon coupling constant of $g_{a\gamma\gamma} < 2.6 \times 10^{-9}$\,GeV$^{-1}$ at a 95\% confidence level.}

\FullConference{Identification of dark matter 2008\\
		 August 18-22, 2008\\
		 Stockholm, Sweden}

\begin{document}

\begin{figure*}
\begin{center}
\includegraphics[width=0.58\textwidth]{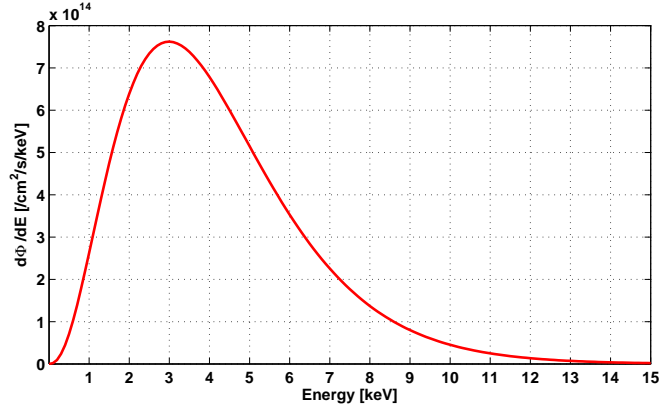}
\caption{Spectral axion flux escaping from the Sun for a coupling constant of $g_{a\gamma\gamma} = 1 \times 10^{-8}$.}
\end{center}
\end{figure*}

\section{Introduction}
The axion has been introduced to explain the strong CP problem arising in QCD. The strong CP problem arises based on null observations of the neutron's electric dipole moment. Spontaneously broken Peccei-Quinn U(1) symmetry leaves a pseudo-Goldstone boson which may be interpreted as a new particle, the axion. The PQ mechanism results in a tiny value of neutron's electric dipole moment which is consistent with experimental constraints \cite{Kim}

Black body photons convert into axions in the presence of the intense electromagnetic field in the Sun's interior. The keV scale energy range of the solar interior creates an axion flux escaping from the Sun in the keV range (Fig.1) \cite{cast2007} : 
\begin{equation}
 \frac{d\Phi}{dE}=6.02 \times 10^{14} cm^{-2} s^{-1} keV^{-1} (g^2_{a\gamma\gamma} GeV^{-2}) E^{2.481}e^{-E/1.205}
\end{equation}
\noindent
Solar axions may be converted back into photons in the strong Coulomb field of an atom by the Primakoff effect \cite{Creswick}.

The 30 Ge and Si crystals operated in the CDMS-II setup are installed in groups of 6 detectors stacked above each other in 5 towers. The details of the detector structure and operational mechanism can be found in \cite{cdmsprd118}.  Each detector is rotated by 60 degrees with respect to the former in the azimuth angle. The [0 0 1] axis is orientated towards the zenith, and the alignment of each crystal's [1 1 0] axis is known to be 0.86$\pm$3 degrees east of true north. This information provides the full knowledge of all crystal axis orientations for each detector. This precise alignment information is unprecedented in experiments searching for solar axions via the Primakoff conversion in crystals, and is crucial to calculating the expected axion conversion rate. 

\begin{figure*}
\begin{center}
\includegraphics[width=0.43\textwidth]{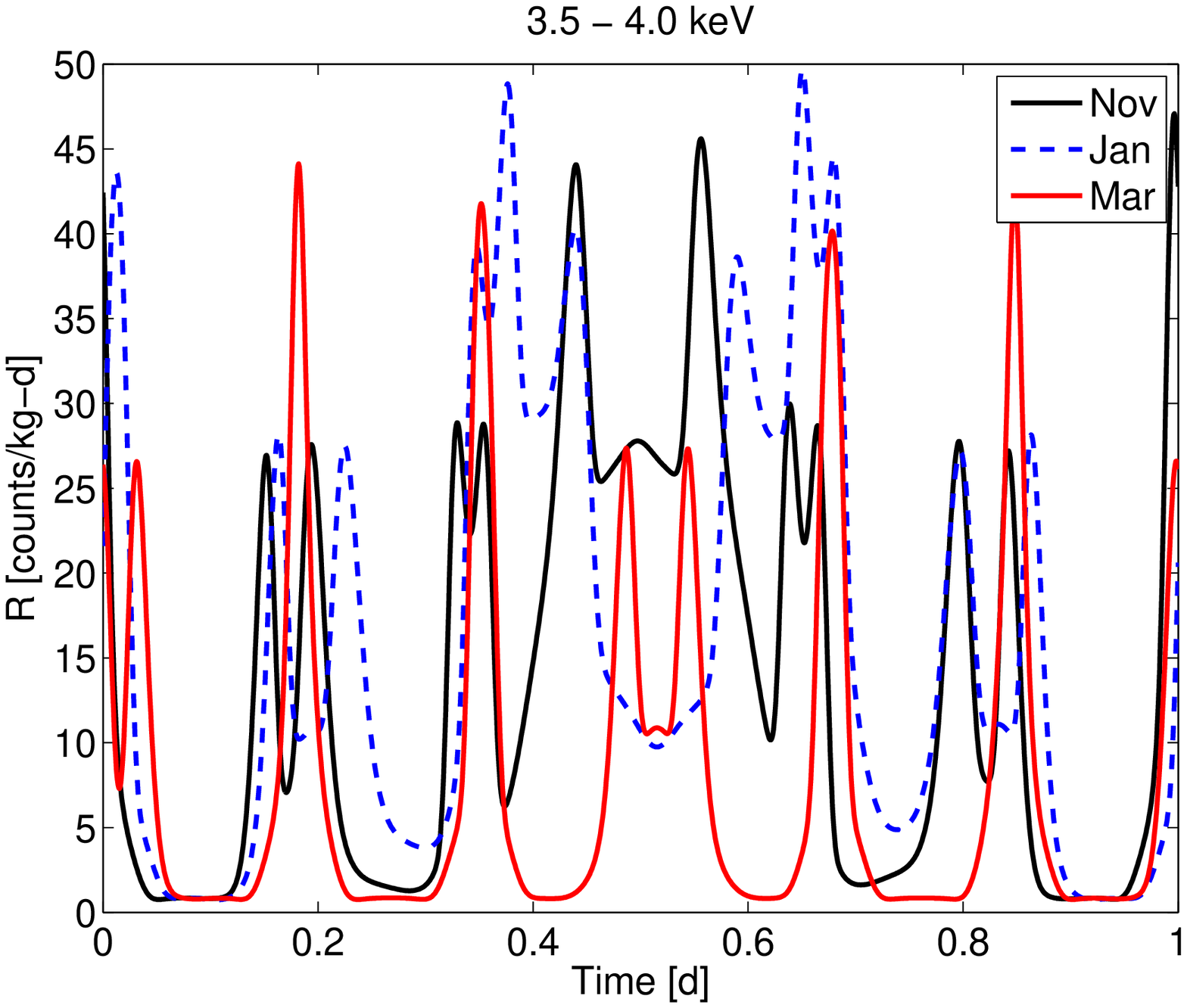}
\includegraphics[width=0.43\textwidth]{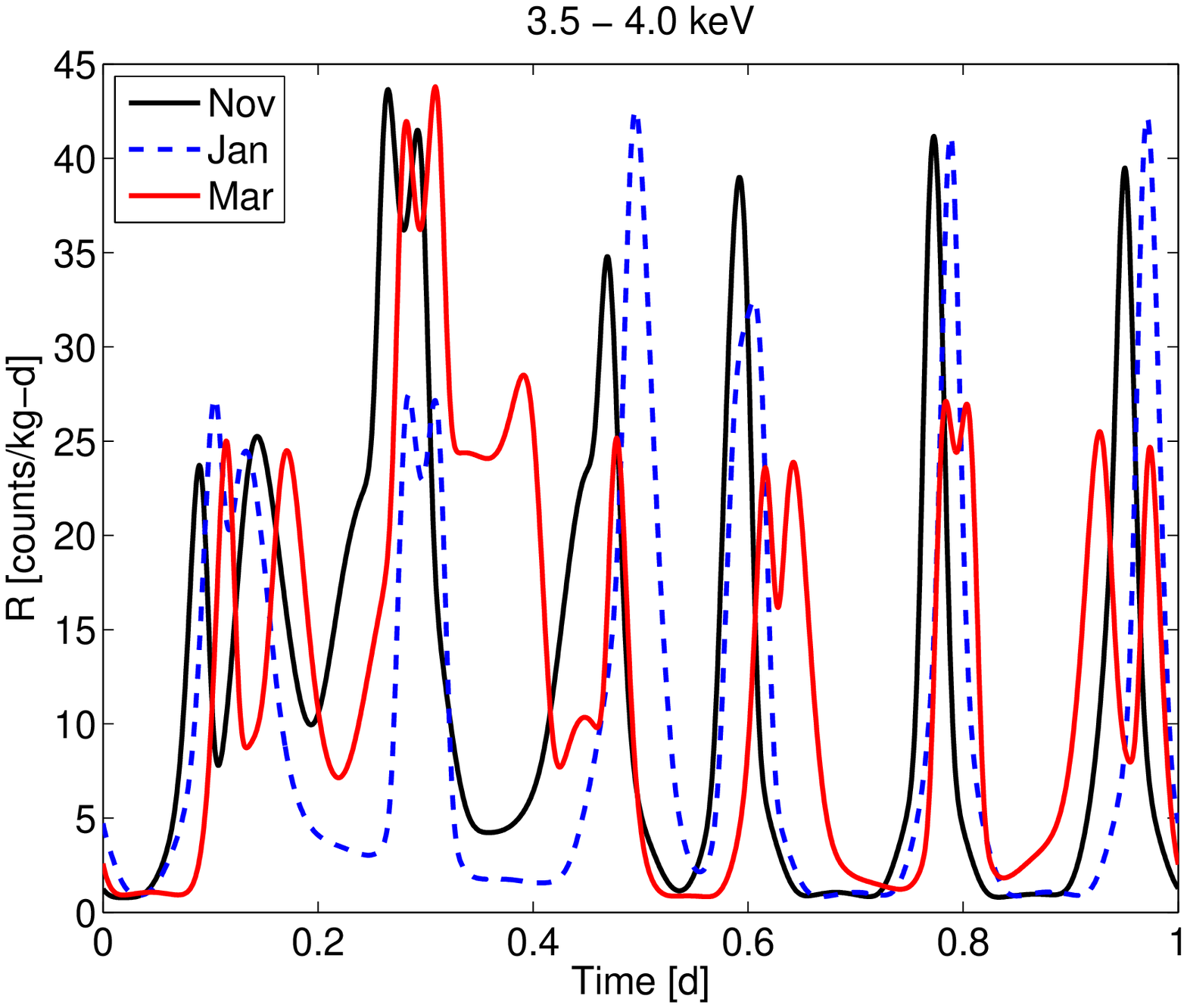}

\caption{Expected conversion spectra in two detectors with different azimuth offset to true north as a function of time. The pronounced variation of the rate with daytime provides powerful background discrimination. Peaks show the reflexes on reciprocal lattice vectors which change their position throughout the year due to the change in the Sun's zenith angle.}
\end{center}
\end{figure*}

\section{Expected conversion spectra}
For very light solar axions the Primakoff process in a periodic crystal lattice is coherent, similar to Bragg reflection of X-rays. This leads to the Bragg condition, namely that the momentum transferred to the crystal must be a reciprocal lattice vector $\overrightarrow{G}$. Integrating over all final photon states the total conversion rate of axions with energy E$_{a}$ in an energy range (E$_{1}$, E$_{2}$) is given by \cite{Creswick,Cebrian} :
\begin{equation}
R(E_1,E_2)[s^{-1}]=(2\pi)^3 2 \hbar c \frac{V}{v_a} \sum_{\overrightarrow{G}} \frac{d\Phi}{dE_a}(E_a) \frac{1}{|\overrightarrow{G}|^2}\frac{g_{a\gamma\gamma}^2}{16\pi^2}|F_a(\overrightarrow{G})S(\overrightarrow{G})|^2sin(2\theta)W
\end{equation}
\begin{equation}
E_a=\hbar c\frac{|\overrightarrow{G}|^2}{2\hat{k}\overrightarrow{G}}
\end{equation}
\noindent
where $V$ is the volume of the crystal, $v_a$ the volume of the elementary cell, $\theta$ is the scattering angle and the function $W$ describes the detector's energy resolution. The direction $\hat{k}$ of the incoming axions is given by the Sun's position at the sky. Thus, the changing position of the Sun throughout a day produces a time dependent conversion rate in the crystals. In addition to the daily variation there is also a seasonal variation due to the changing zenith angle of the Sun in one year. The variation of the counting rate with daytime is shown in Fig 2. for two groups of detectors used in this analysis. The alternating orientations of the individual crystals in the experimental setup provide different patterns of solar axion conversion, making a false positive result extremely unlikely.

\begin{figure*}
\begin{center}
\includegraphics[width=0.46\textwidth]{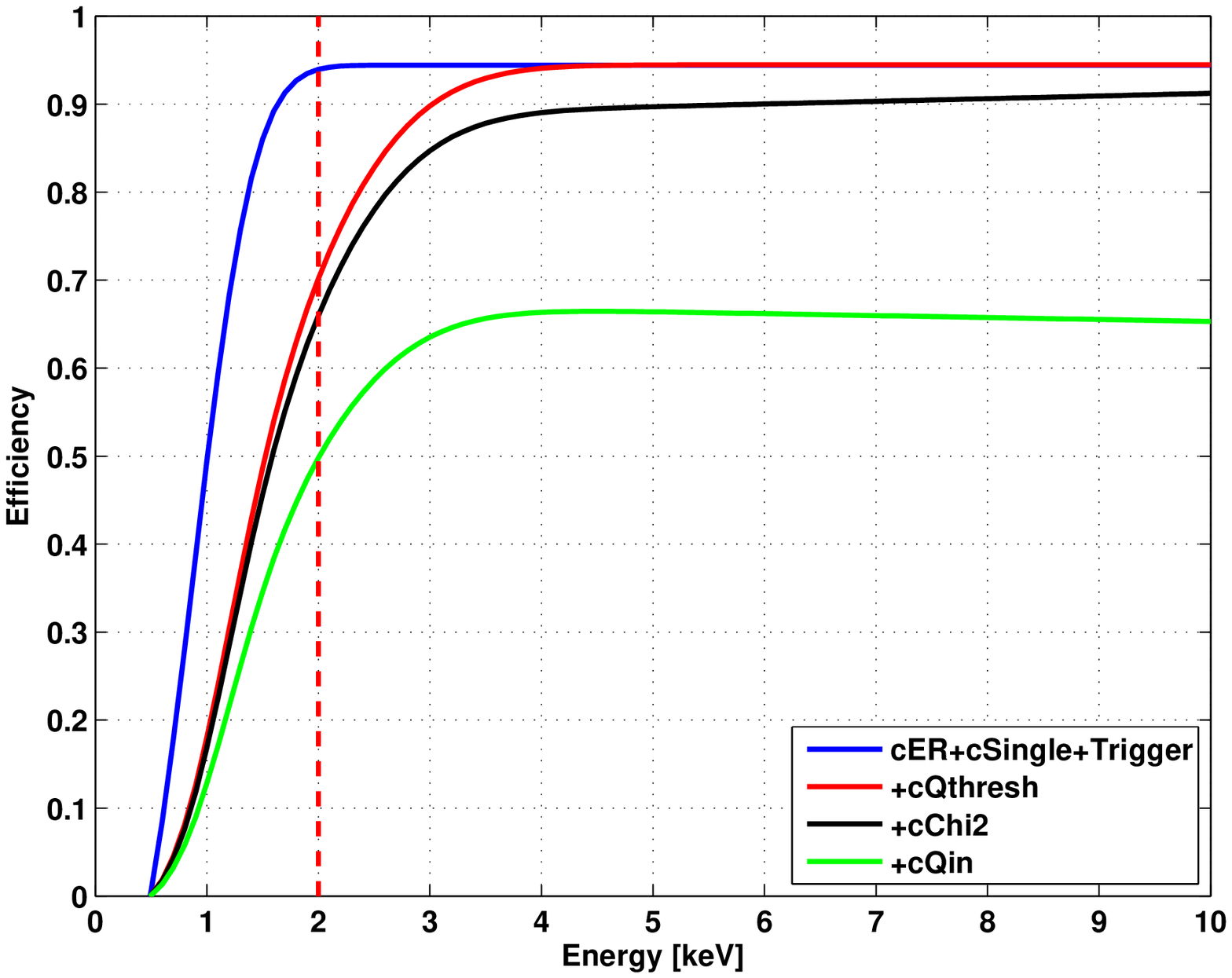}
\includegraphics[width=0.46\textwidth]{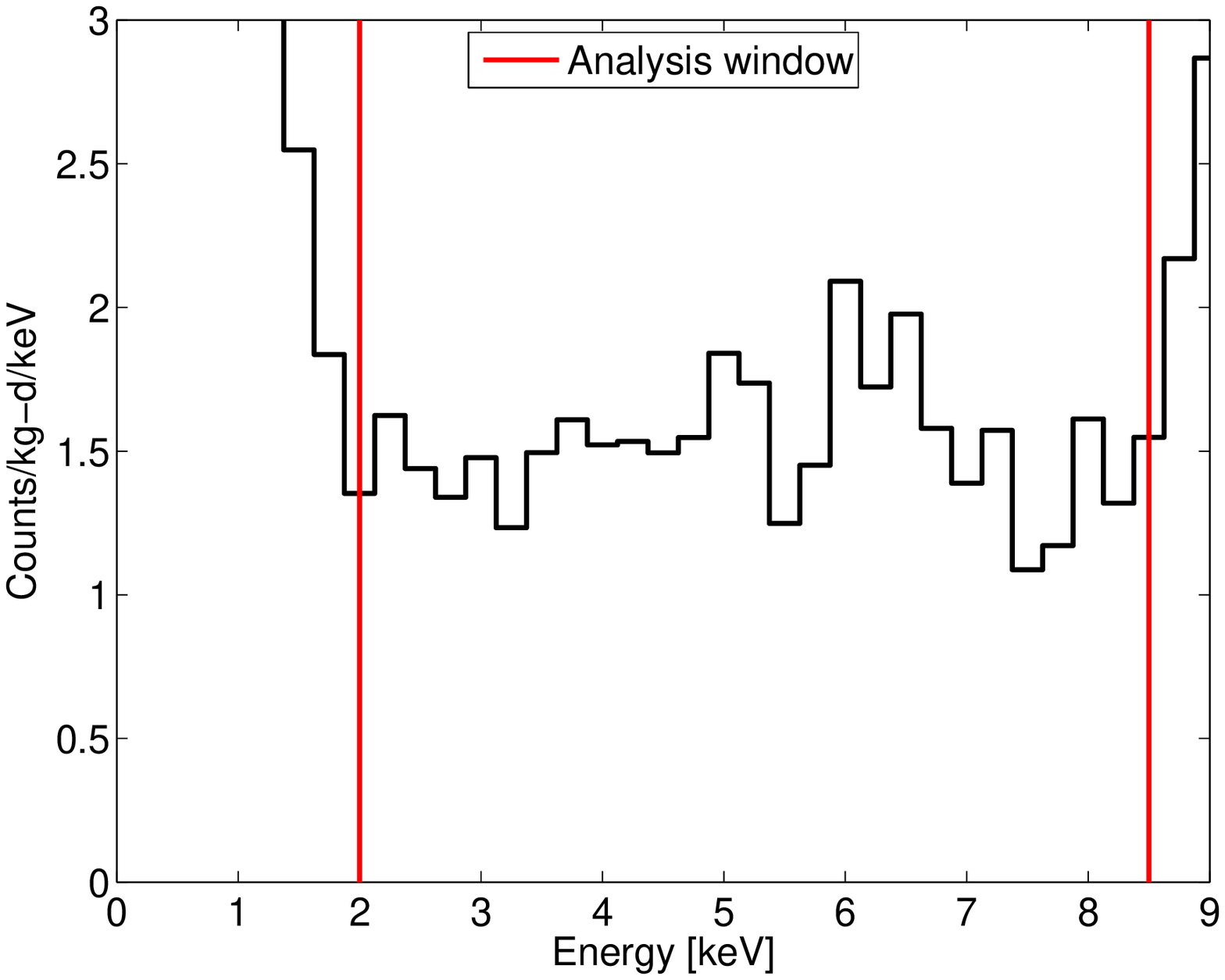}

\caption{Left panel: Detection efficiency as a function of energy after selecting single (cSingle), electron recoil (cER) events, passing noise rejection (cQthresh) and good charge pulse identification (cChi2) cuts which are in the fiducial volume (cQin). Right panel: Efficiency corrected, co-added low energy spectra of the 13 Ge detectors considered in this analysis.}
\end{center}
\end{figure*}

\section{Analysis}
The conversion of axions to photons with energies $\leq$ 15 keV in the crystals requires a good energy resolution in this energy range for a high signal to noise ratio.
Neutron capture on $^{70}$Ge produces $^{71}$Ge during $^{252}$Cf calibrations of the detectors. The de-excitation of $^{71}$Ga, following the electron capture decay of $^{71}$Ge, causes 10.36 keV electron recoil events. The typical energy resolution of the Ge detectors at 10.36 keV is better than 3\%. The functional form of the energy resolution for each detector is determined by interpolating the energy resolution at 10.36 keV to the zero-energy resolution, given by the width of the noise distribution.

Since the CDMS-II detectors are capable to distinguish between electron and nuclear recoils we only select electron recoils in this analysis, which are expected from the conversion photons. Candidate events must also be single scatters within the fiducial volume satisfying data quality and noise rejection cuts. Combining these requirements gives a typical detection efficiency of 50-60\% in the energy range of 2-10 keV (Fig. 3, left panel). For the 13 detectors considered in this analysis (three more may be added in future analyzes), the co-added, efficiency corrected background rate is $\sim$ 1.5 counts/kg/day/keV. The analysis window, defined from 2- 8.5 keV is determined by the expected axion flux, background rate and detection efficiency (Fig. 3, right panel).

To extract a possible axion signal from the data we perform an unbinned likelihood maximization. We express the event rate per unit, measured energy (E), per unit time (t) and per detector (d) of a solar axion signal with background as: 
\begin{equation}
 R(E,t,d)=\lambda A(E,t,d)+B(E,d)
\end{equation}
\noindent
where $A(E,t,d)$ is the expected event rate for a coupling constant of $g_{a\gamma\gamma}=10^{-8}$ GeV$^{-1}$ which includes the detector specific energy resolution and detection efficiency $\varepsilon(E,d)$. The scaling factor for the actual value of $g_{a\gamma\gamma}$ is denoted by $\lambda$. The background $B(E,d)$ is approximated by
\begin{equation}
 B(E,d)=\varepsilon(E,d)(C(d)+D(d)E+H(d)/E)
\end{equation}
\noindent
where $C(d)$, $D(d)$ and $H(d)$ are free parameters. The fitting is done by maximizing the unbinned log likelihood with respect to $\lambda$ and the background parameters, for individual events {\it i} ,
\begin{equation}
log(\mathcal{L}) = -R_T + \sum_i log(R(E_i,t_i,d_i))
\end{equation}
where $R_T$ is the total sum of the event rate ($R$) over energy, time and detectors.

The best estimator for the scaling factor $\lambda = (g_{a\gamma\gamma} \times 10^8 GeV)^4 $ from the maximization is $\lambda=(0.09 \pm 2.2) \times 10^{-3}$.  The error is determined from the value at which the profile likelihood has decreased by 0.5 from the maximum value (Fig. 4, right panel). Since the best estimator is compatible with zero, the upper limit on the scaling factor at a 95\% CL has been determined by integrating the profile likelihood in the physical allowed region ($\lambda > 0$).

\begin{figure*}
\begin{center}
\includegraphics[width=0.46\textwidth]{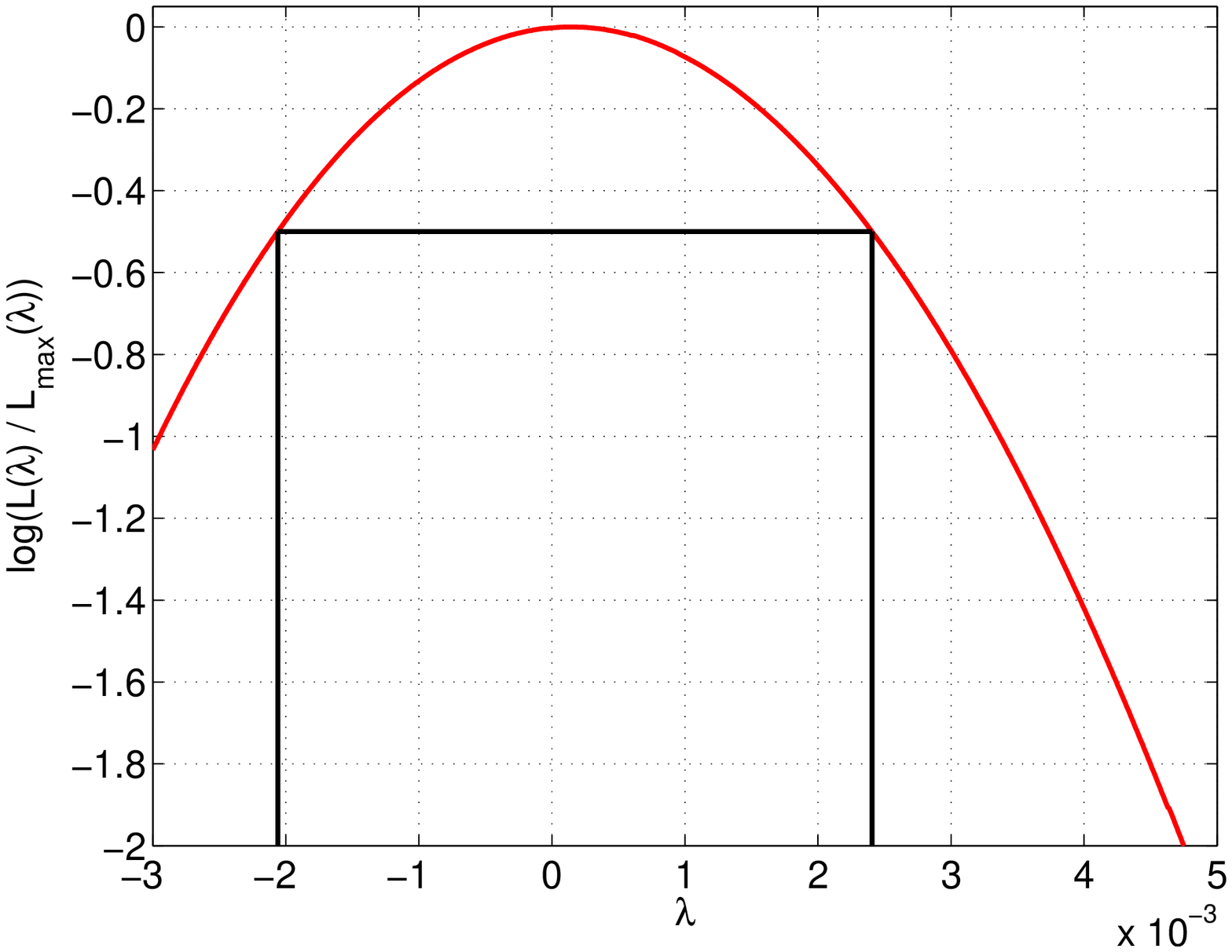}
\includegraphics[width=0.50\textwidth]{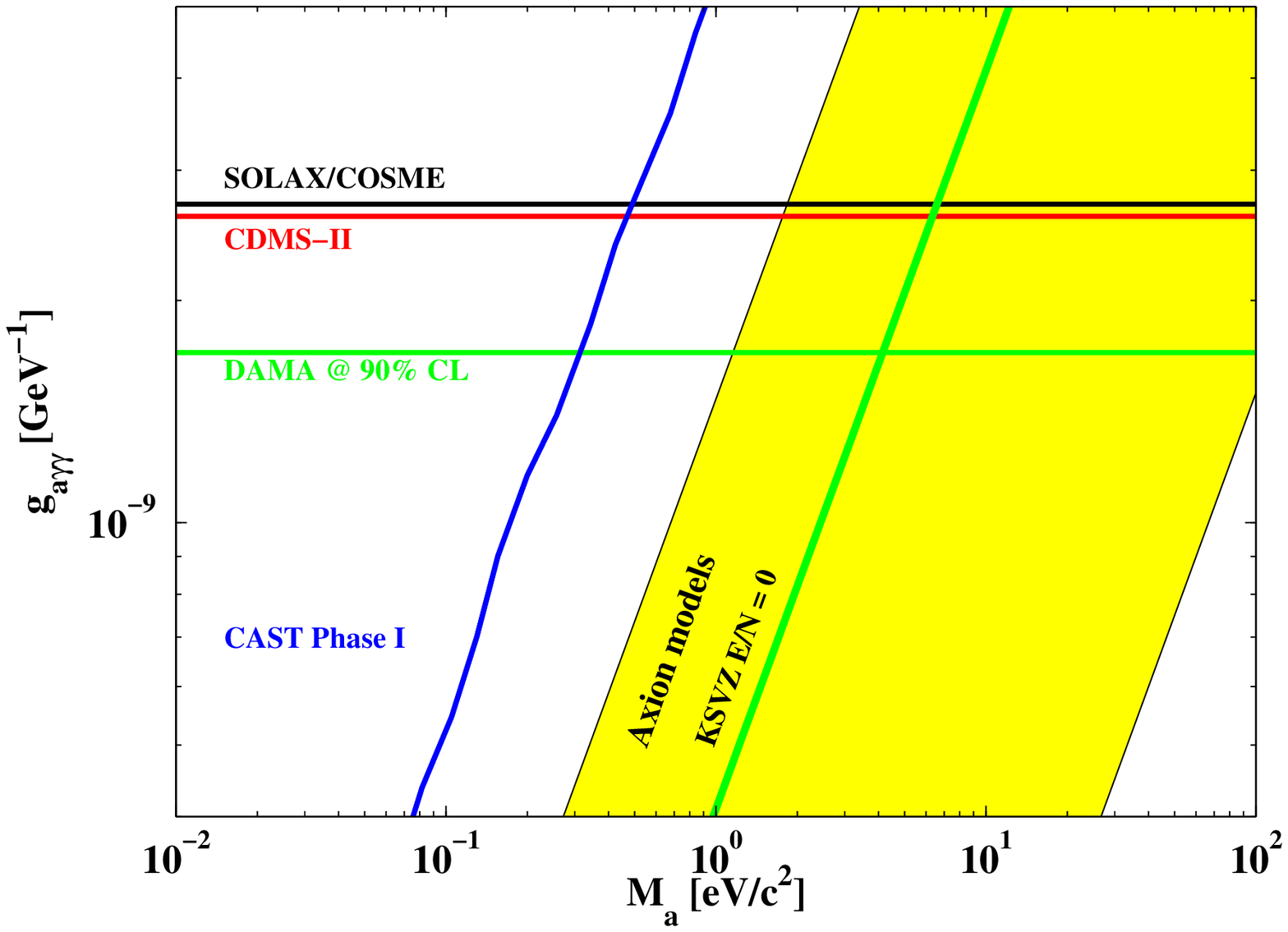}

\caption{Left panel: Profile likelihood of the scaling factor $\lambda$. The black/solid lines mark the points at which the profile likelihood has decreased by a factor of 0.5 from it's maximum. Right panel: Upper limits on $g_{a\gamma\gamma}$ from several experiments. The upper limit from this analysis is shown in red. }
\end{center}
\end{figure*}

\section{Results}
The upper limit at a 95\% CL on the photon axion coupling constant from this analysis is:
\begin{equation}
 g_{a\gamma\gamma} < 2.6 \times 10^{-9} GeV^{-1}
\end{equation}
\noindent
This result lies below the current best limit set by germanium detectors from the SOLAX experiment, although the exposure of this analysis (289 kg-days) is only $\sim$ 40\% of the SOLAX \cite{Solax} exposure. The advantages  of the CDMS-II experiment are the low background rate and the precise knowledge of the crystal orientations. The result of this analysis is compared to other experimental results in Fig. 4 (right panel). This analysis inspires the prospect that future large crystal detector arrays such as SuperCDMS and GERDA may provide competitive sensitivity on the photon-axion coupling constant for $g_{a\gamma\gamma} < 10^{-9}$ in the high mass region (< 1keV) not easily accessible to helioscopes such as the CAST \cite{CASTexp} experiment.


\begin{thebibliography}{99}
\bibitem{Kim} Kim J. E., Physics Reports 150, Nos. 1\&2, 1-177 (1987). 
\bibitem{cast2007} Andriamonje S. {\it et al.} (CAST Collaboration), J. Cosmol. Astropart. Phys. 04, 010 (2007). 
\bibitem{Creswick} Creswick R.~J. {\it et al.}, Phys. Lett. B, 427, 235 (1998).
\bibitem{cdmsprd118} Akerib D. S. {\it et al.} (CDMS Collaboration), Phys, Rev. D 72, 052009 (2005).
\bibitem{Cebrian} Cebri\'{a}n S. {\it et al.}, Astropart. Phys. 10, 397 (1999)
\bibitem{Solax} Avignone F.~T. {\it et al.} (SOLAX Collaboration), Phys. Rev. Lett. 81, 5068 (1998).
\bibitem{CASTexp} Zioutas K. {\it et al.} (CAST Collaboration), Phys. Rev. Lett. 94, 121301 (2005).

\end{thebibliography}
\end{document}